\documentclass{jps-cp}
\usepackage{txfonts} 

\title{Numerical Study of S=1/2 Heisenberg Antiferromagnet \\
on the Floret Pentagonal Lattice}

\author{Rito \textsc{Furuchi}$^{1}$, Hiroki \textsc{Nakano}$^{1}$, and T$\hat{\rm{o}}$ru \textsc{Sakai}$^{1,2}$}

\inst{$^{1}$ Graduate School of Science, University of Hyogo, Kamigori, Hyogo 678-1297, Japan\\
$^{2}$ National Institutes for Quantum and Radiological Science and Technology,
SPring-8 Sayo, Hyogo 679-5148, Japan\\
}

\email{dribloodwolf@yahoo.co.jp}

\recdate{July 14, 2022}

\abst{The $S=1/2$ Heisenberg antiferromagnet on
the floret-pentagonal lattice with
two kinds of interaction strength
is studied
by the numerical-diagonalization method.
It is known that,
near the five-ninth
of the saturation magnetization,
this system shows a magnetization jump
that is not accompanied by magnetization plateaux.
We focus our attention
on the behavior of this system around the five-ninth
of the saturation magnetization;
the changes of the magnetization jump and plateau at and around this
magnetization
are clarified from the diagonalization data for finite-size systems
up to 45 sites.
}

\kword{$S=1/2$ Heisenberg antiferromagnet,The floret-pentagonal lattice,
Magnetization process, Numerical diagonalization}

\begin{document}
\maketitle

\section{Introduction}
Frustration in magnetic materials has attracted much attention
of many condensed-matter physicists.
The frustration is typically caused by a local structure
of antiferromagnetic interactions forming
a polygon with an odd number of sides
\---
an odd-gon
\---
.
The simplest case is the triangular structure; magnetism
on various lattices including such triangular structure, for example,
the triangular lattice and the kagome lattice,
have been extensively and intensively studied.
The next possibility for the
odd-gon
is a pentagon.
However, a relatively much smaller number of investigations
have been carried out for the magnetism of a system on a pentagonal lattice.
As such a two-dimensional pentagonal lattice,
the Cairo-pentagonal-lattice Heisenberg antiferromagnet was studied
and the system shows the characteristic magnetization process
with magnetization plateaux and jumps\cite{Rousochatzakis_Cairo,
HNakano_Cairo_lt,Isoda_Cairo_full}.
Note here that candidate materials for the Cairo-pentagonal-lattice
antiferromagnet were studied recently\cite{Abakumov_PRB2013,
Tsirlin_PRB2017,Chattopadhyay_SciRep2017,Cumby_DaltonTrans2016,
Beauvois_PRL2020}.
As systems including local pentagonal structure,
studies concerning
spherical kagome cluster\cite{Fukumoto_spherical_kagome_PTEP2014},
dodecahedral cluster\cite{Konstantinidis_dodeca_PRB2005,
Konstantinidis_dodeca_JPCM2016}, and
icosidodecahedron cluster
\cite{Exler_icosidodecahedron_PRB2003,Schroder_icosidodecahedron_PRB2005}
are also known.

Recently, the floret-pentagonal-lattice (FPL) antiferromagnet was
investigated\cite{Furuchi_JPCO2021}
as the second two-dimensional case among pentagonal lattices.
The FPL Heisenberg antferromagnet shows magnetization plateaux
in its magnetization process at one-ninth, one-third,
and seven-ninth
of the saturation magnetization.
The magnetization plateaux are related to the number of spin sites
in each unit cell of this lattice, namely, nine.
Spin sites of the FPL are divided into two groups;
a coordination number of sites on one group is six and
that on the other group is three.
Let us
consider antiferromagnetic interaction
bonding a site of coordination number to be six ($J_{1}$) and
antiferromagnetic interaction not bonding such sites ($J_{2}$).
Reference~\ref{Furuchi_JPCO2021} reported that the magnetization plateaux
at one-third and seven-ninth
of the saturation magnetization
get smaller and close when $J_2/J_1$ is increased.

Let us we focus our attention
on the behavior of the FPL antiferromagnet
at and near the five-ninth
of the saturation magnetization.
In the uniform case, the FPL antiferromagnet does not show
a magnetization plateau at this
magnetization;
on the other hand,
the system shows a peculiar magnetization jump near this
magnetization.
Under these circumstances,
the purpose of this study is to clarify the behavior near this
magnetization
during the variation of $J_2/J_1$ by the Lanczos-diagonalization method
for finite-size clusters of this system.
We successfully capture the appearance the magnetization jump
near this
magnetization
in a specific range of $J_2/J_1$.
We also find that the magnetization plateau at five-ninth
of the saturation appears for $J_2/J_1$ that is larger
than a specific value of this ratio.

This paper is organized as follows.
In the next section, the model Hamiltonian will be introduced.
The method of calculations will also be explained.
The third section is devoted
to the presentation and discussion of our results
concerning the magnetization jump near the five-ninth
of the saturation.
In the fourth section, we discuss the appearance of the magnetization
plateau at this
magnetization.
In the final section,
a summary of our results and some remarks will be given.

\section{Model and calculation}
\begin{figure}[tbp]
 \begin{minipage}{0.5\hsize}
  \begin{center}
    \includegraphics[width=70mm]{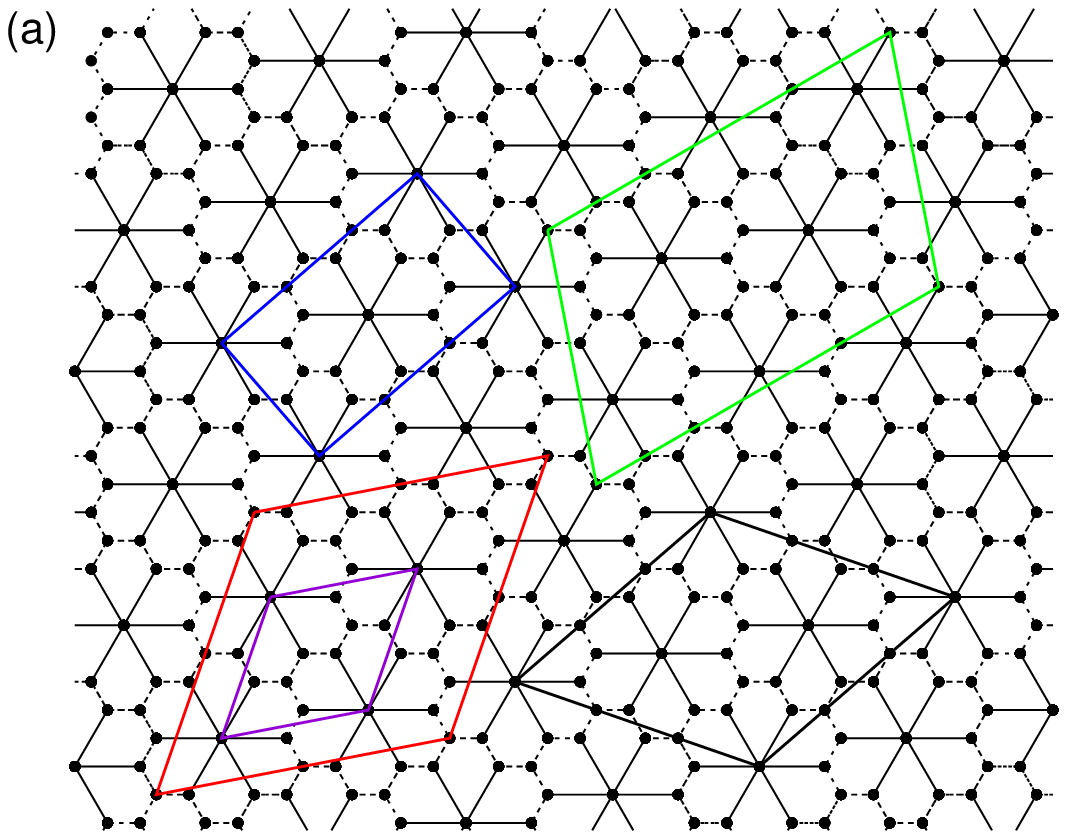}
  \end{center}
 \end{minipage}
 \begin{minipage}{0.5\hsize}
  \begin{center}
    \includegraphics[width=70mm]{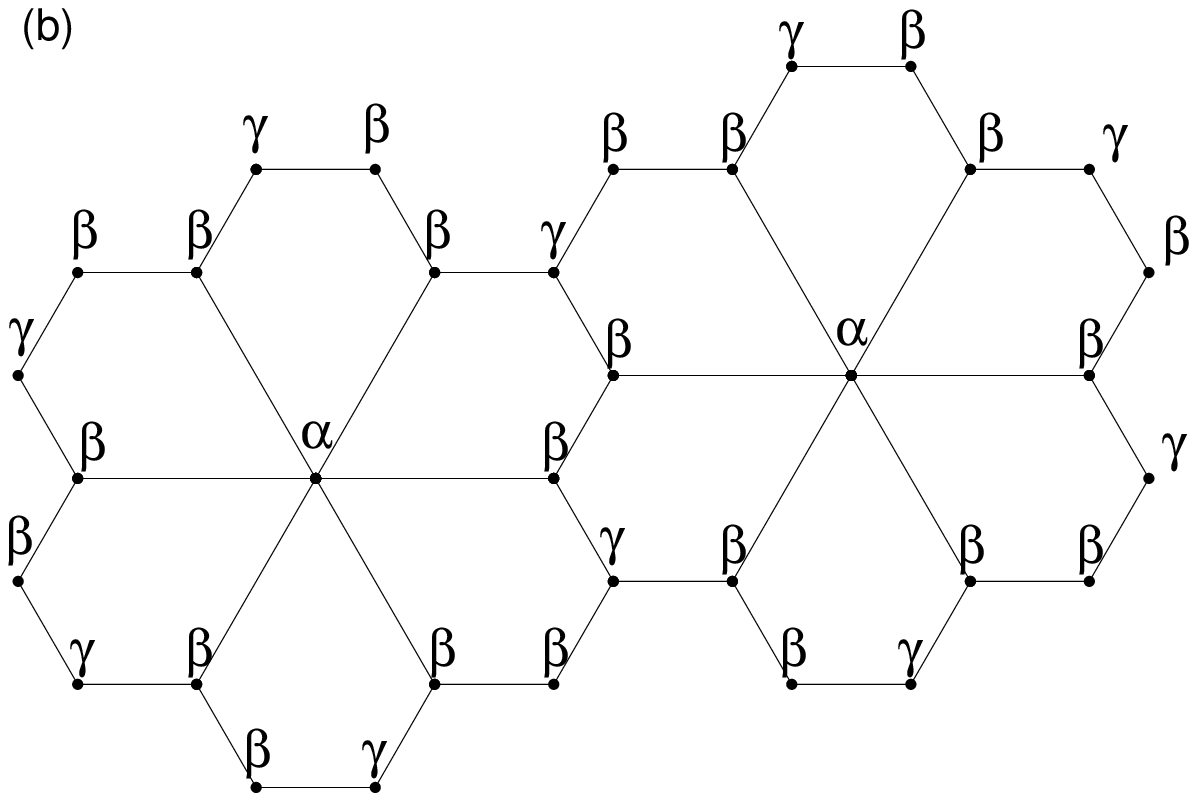}
  \end{center}
 \end{minipage}
 \caption{Panel (a) shows the floret-pentagonal lattice and
finite-size clusters treated in this study.
Solid and broken lines denote interaction with amplitude $J_1$ and $J_2$,
respectively.
Violet, blue, black, red, and green solid lines
correspond to the finite-size clusters
for $N=9$, $18$, $27$, $36$, and $45$, respectively.
Panel (b) shows groups of vertices: $\alpha$, $\beta$, and $\gamma$.}
 \label{lattice}
\end{figure}

We investigate the $S=1/2$ Heisenberg antiferromagnet on the FPL.
The Hamiltonian of the investigated model is described by
${\cal H} = {\cal H}_{0} + {\cal H}_{\mathrm{zeeman}}$
where
\begin{eqnarray}
  {\cal H}_{0}&=&\sum_{\rm{solid \hspace{2pt} bonds}}
  J_{1} \mbox{\boldmath $S$}_{i} \cdot \mbox{\boldmath $S$}_{j}
  +\sum_{\rm{dotted \hspace{2pt} bonds}}
  J_{2} \mbox{\boldmath $S$}_{i} \cdot \mbox{\boldmath $S$}_{j}, \
  {\cal H}_{\rm zeeman} = - h \sum_{j} S_{j}^{z} .\label{spinhamil}
\end{eqnarray}
Here, $\mbox{\boldmath $S$}_{i}$ denotes the $S = 1/2$ spin operator
at site $i$ illustrated by the closed circle
at a vertex in Fig.~\ref{lattice}(a).
In the first term of Eq.~\eqref{spinhamil},
the sum is taken for the bonds depicted by the solid line
in Fig.~\ref{lattice}(a), and in the second term,
the sum is taken for the bonds depicted by the dotted line.
Energies are measured in units of $J_{1}$ in Eq.~\eqref{spinhamil}.
Since we examine the case of antiferromagnetic interaction,
we put $J_{1}=1$, $J_{2}>0$ hereafter.
We define $\eta=J_{2}/J_{1}$ as the ratio of these interactions.
In the case $\eta=1$, all interactions are equivalent,
when the behavior of this system was studied in Ref.~\ref{Furuchi_JPCO2021}.
In that case,
some nontrivial phenomena due to frustration were reported.
In the present study,
we investigate what happens when $\eta$ is varied
concerning the phenomena that were observed in Ref.~\ref{Furuchi_JPCO2021}.

Figure~\ref{lattice}(a) also shows the shape of finite size clusters
treated in the present study.
The number of spin sites is denoted as $N$.
Note that
the finite-size clusters for $N=9$, $27$, and $36$ are rhombic,
whereas those for $N=18$ and $45$ are not rhombic.
Although these nonrhombic clusters therefore show symmetries that are
different from the FPL, calculations of $N=18$ and $45$
contribute to deepen our understanding of the FPL antiferromagnet.
In all the finite-size clusters, the periodic boundary condition is employed.
Note also that the magnetization process for a 45-site cluster
was first reported in the case
of the kagome-lattice antiferromagnet\cite{HNakano_JPSJ2018kagome45}
and that the present study for the FPL antiferromagnet
is the second report of a 45-site magnetization process
to the best of our knowledge.

The FPL has originally been known
in the tiling problem\cite{Schattschneider1978}.
Figure~\ref{lattice}(b) shows the grouping of vertices of this lattice.
A unit cell of the FPL contains nine vertices,
which are divided into two groups at first.
One is a group of vertices of the type characterized
by the coordination number $z=6$ and
the other group consists of vertices with $z=3$.
The former vertex is called $\alpha$ sites,
and the latter is further divided into two groups:
those linked by a bond with $\alpha$ are called $\beta$ sites,
and those not linked are called $\gamma$ sites.

In this study, the
ground-state
energy of ${\cal H}_{0}$ is calculated
in the subspace characterized by $M$ defined by $\sum_{j} S_{j}^{z}$.
The calculation to obtain the energy by diagonalizing ${\cal H}_{0}$
is based on the Lanczos algorithm and/or the Householder algorithm.
The energy is denoted by $E(N,M)$.
The saturation value of $M$ is defined as $M_{\rm s}(=SN)$,
until which $M$ increases discretely with $\delta M=1$.
The magnetization process is determined so that the magnetization increases
from $M$ to $M+1$ at magnetic field $h=E(N,M+1)-E(N,M)$
when the magnetization monotonically increases with $h$.
When the monotonic increase disappears, there appears a jump;
the Maxwell construction should be carried out to capture
the behavior of the jump.
We evaluate local magnetization $m_{\rm loc}$
defined as $(1/N_{\xi})\sum_{j\in \xi}\langle S_{j}^{z}\rangle$,
where $\xi$ takes $\alpha$, $\beta$, and $\gamma$;
$\langle{\cal O}\rangle$ represents the expectation value of an operator
${\cal O}$ with respect to the lowest-energy state
within the subspace with a fixed $M$ of interest.
For simplicity, here, we define $m=M/M_{\rm{s}}$ as the normalized magnetization.
Part of Lanczos diagonalizations has been performed
using the MPI-parallelized code, which was originally developed
in the research of the Haldane gaps\cite{nakanoterai}.
The usefulness of our program was demonstrated
in several large-scale parallelized calculations\cite{HNakano_JPSJ2011,
HNakano_JPSJ2018,HNakano_JPSJ2019,HNakano_S1HaldaneGap_JPSJ2022}.

\section{Magnetization jump}
\begin{figure}[tbp]
 \begin{minipage}{0.5\hsize}
  \begin{center}
   \includegraphics[width=70mm]{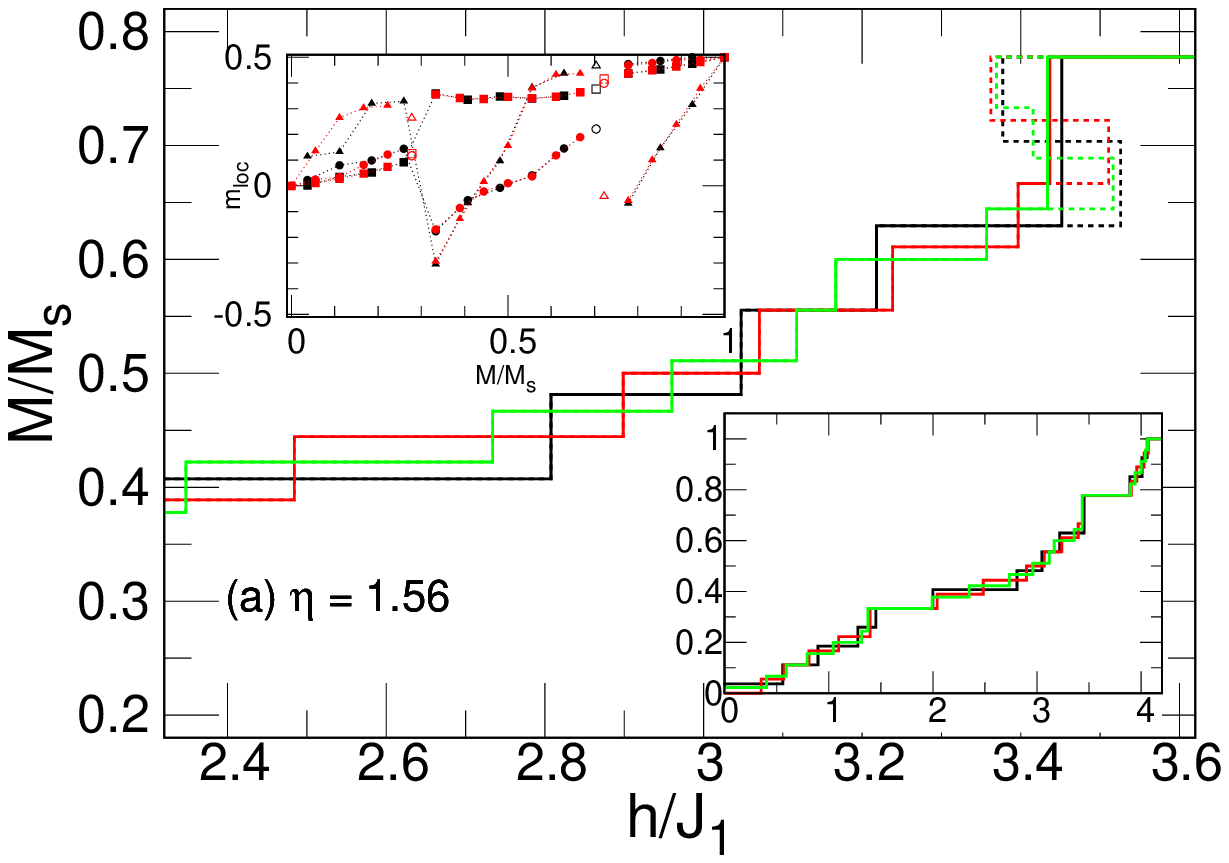}
  \end{center}
 \end{minipage}
 \begin{minipage}{0.5\hsize}
  \begin{center}
   \includegraphics[width=70mm]{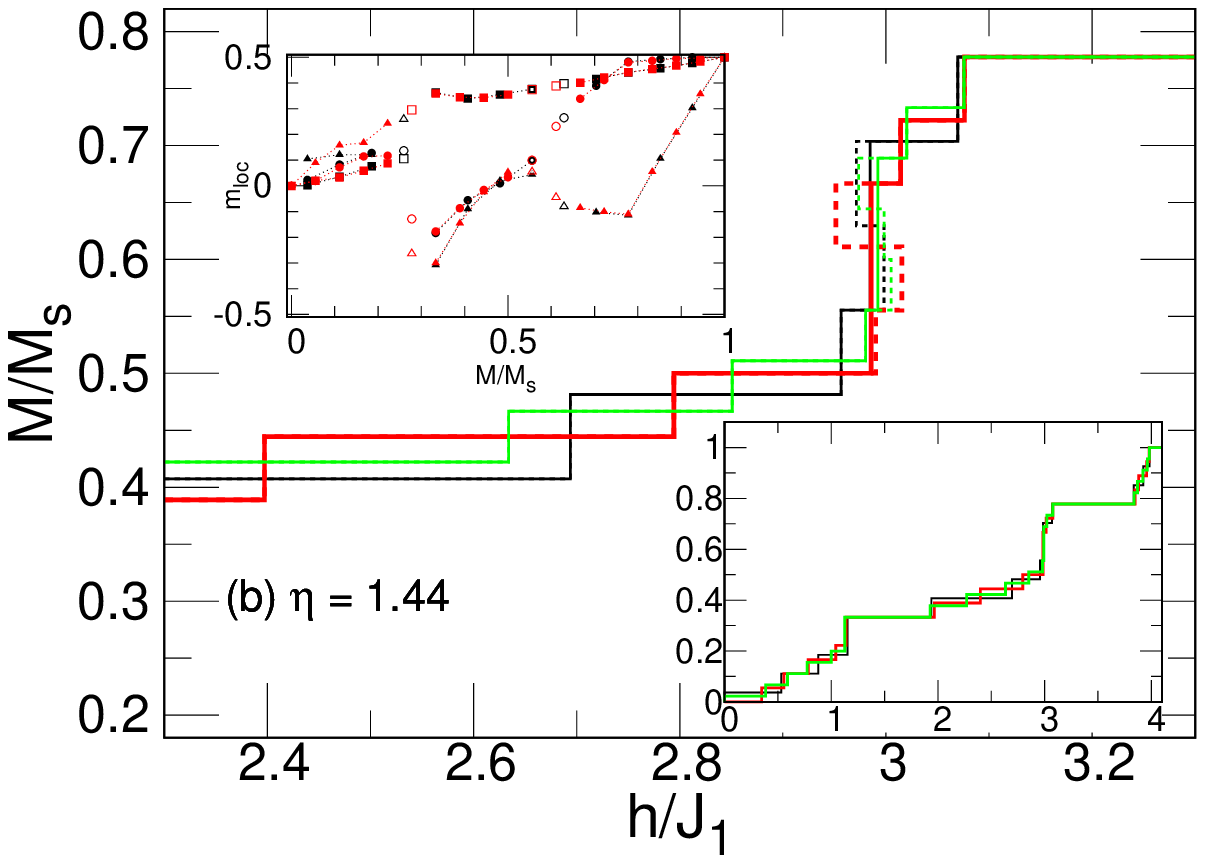}
  \end{center}
\end{minipage}
 \caption{Panels (a) and (b) show the magnetization process and
local magnetization for $\eta=1.56$ and $1.44$, respectively.
The green, red, and black lines and dots represent
results for $N=45$, $36$, and $27$ systems, respectively.
The main panel shows a magnified view around the magnetization jump;
the solid lines depict the magnetization process
after the Maxwell construction is applied.
The dashed lines depict the results
before the Maxwell construction is applied,
which explicitly show unrealized states.
The lower-right inset depicts the magnetization process in the entire range
up to the saturation.
The upper-left inset shows the averaged local magnetization in each system
for the rhombic clusters for $N=36$ and 27.
The averaging is carried out among each of groups,
$\alpha$, $\beta$, and $\gamma$ sites that are represented by
triangles, squares, and circles, respectively.
Open symbols indicate skipped states
because these states do not become a ground state
even under any magnetic fields
although closed symbols indicate realized states.
}
 \label{magpro}
\end{figure}
Recall here that in the system with $\eta=1$,
a jump for the $N=36$ cluster appears as a
skipped case
of $m=11/18$\cite{Furuchi_JPCO2021}.
This jump is significantly different from jumps that have been reported
in many other frustrate systems.
Such magnetization jumps are associated with a specific magnetization plateau.
However, the jump of the $N=36$ FPL-antiferromagnetic cluster
at $m=11/18$ 
appears
away from both the plateaux at $m=1/3$ and 7/9.
In this section,
we
clarify what happens in the magnetization process of this model
by observing the change of this jump when $\eta$ is varied.

Now, let us consider the cases for larger $\eta$;
results of magnetization processes for $N=27$, $36$, and $45$
are depicted in Fig.~\ref{magpro} for $\eta=1.56$ and 1.44
in panel (a) and (b), respectively.
Figure~\ref{magpro} 
presents
the magnetization processes
in the entire range in the lower-right inset and
a zoom-in view near the jump
in the main panel.
Let us examine the change of the magnetization process
when $\eta$ is decreased.
Recall before observing Fig.~\ref{magpro} that
the $m=7/9$ plateau begins to open
around $\eta\sim 1.6$\cite{Furuchi_JPCO2021}.
For $\eta=1.56$, there appears
a magnetization jump at the lower-edge of the $m=7/9$ plateau
in Fig.~\ref{magpro}(a).
For $\eta=1.44$, next,
the magnetization jump departs from the $m=7/9$ plateau.

Next, let us observe the relationship between
the change of the magnetization jump
and averaged local magnetizations $m_{\rm loc}$
in the upper-left insets of Fig.~\ref{magpro}.
In particular, a marked behavior appears
in the results of $m_{\rm loc}$ for $\alpha$ sites.
For $\eta=1.56$, the clear discontinuous behavior of $m_{\rm loc}$
in $\alpha$-site results is observed
between data at $m=7/9$ and those for $0.6 < m < 0.7$.
For $\eta=1.44$, on the other hand,
states for $m\sim 0.6$ actually are realized as
the ground states under a specific magnetic field.
The results of $m_{\rm loc}$ of $\alpha$ sites in these states
and $m_{\rm loc}$ at $m=7/9$ show a continuous behavior.
At the same time, there still exists
a discontinuous behavior across the jump around $m\sim 0.6$.
On the other hand,
there are no significant changes in the spin states around $m\sim 0.5$.

From observing the behavior of this system in Fig.~\ref{magpro},
the jump originally begins to appear in an association
with the opening of the plateau at $m=7/9$
when $\eta$ is decreased from a large-$\eta$ side.
In even smaller $\eta$,
there appear new states between the jump and the $m=7/9$ plateau
under the situation that the jump still survives.
Therefore, the new states
play an essential role in the formation of the magnetization jump
away from any magnetization plateaux.

\begin{figure}[tbp]
  \begin{minipage}{0.5\hsize}
    \begin{center}
      \includegraphics[width=70mm]{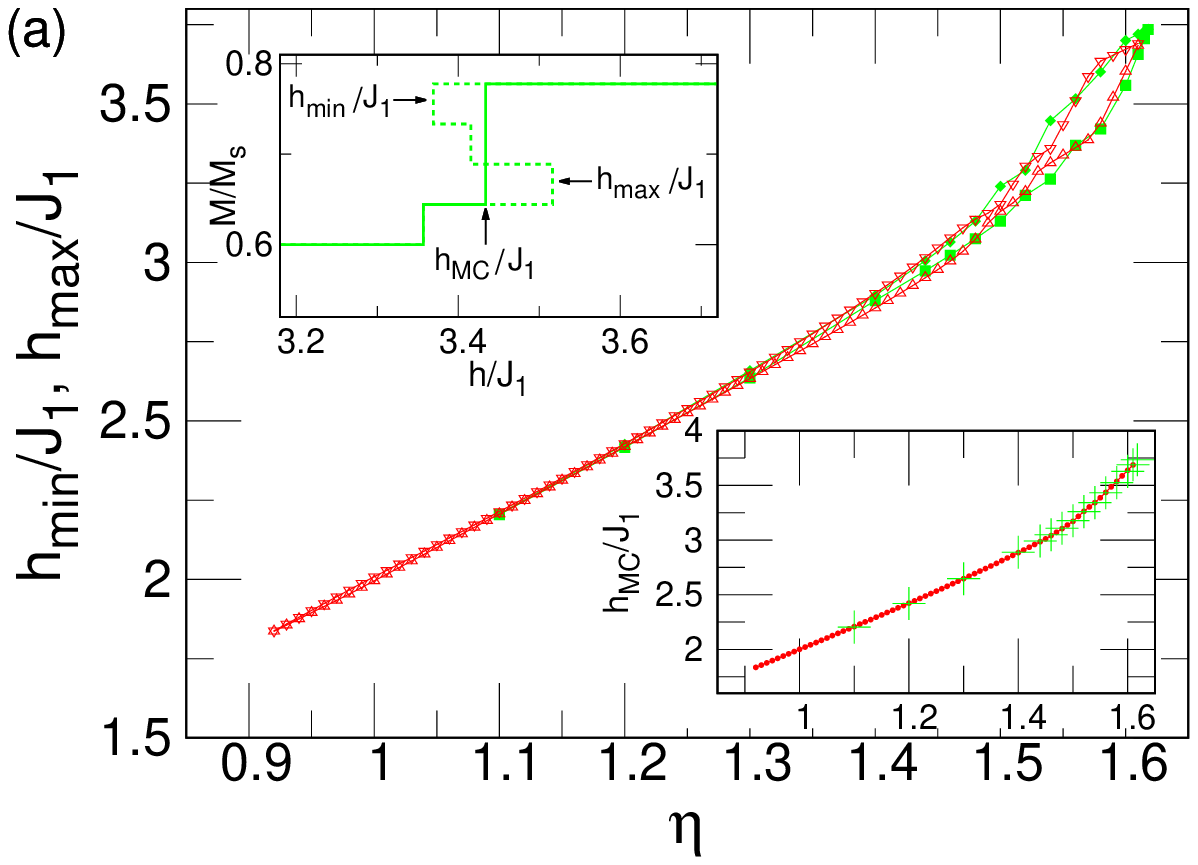}
    \end{center}
  \end{minipage}
 \begin{minipage}{0.5\hsize}
   \begin{center}
      \includegraphics[width=70mm]{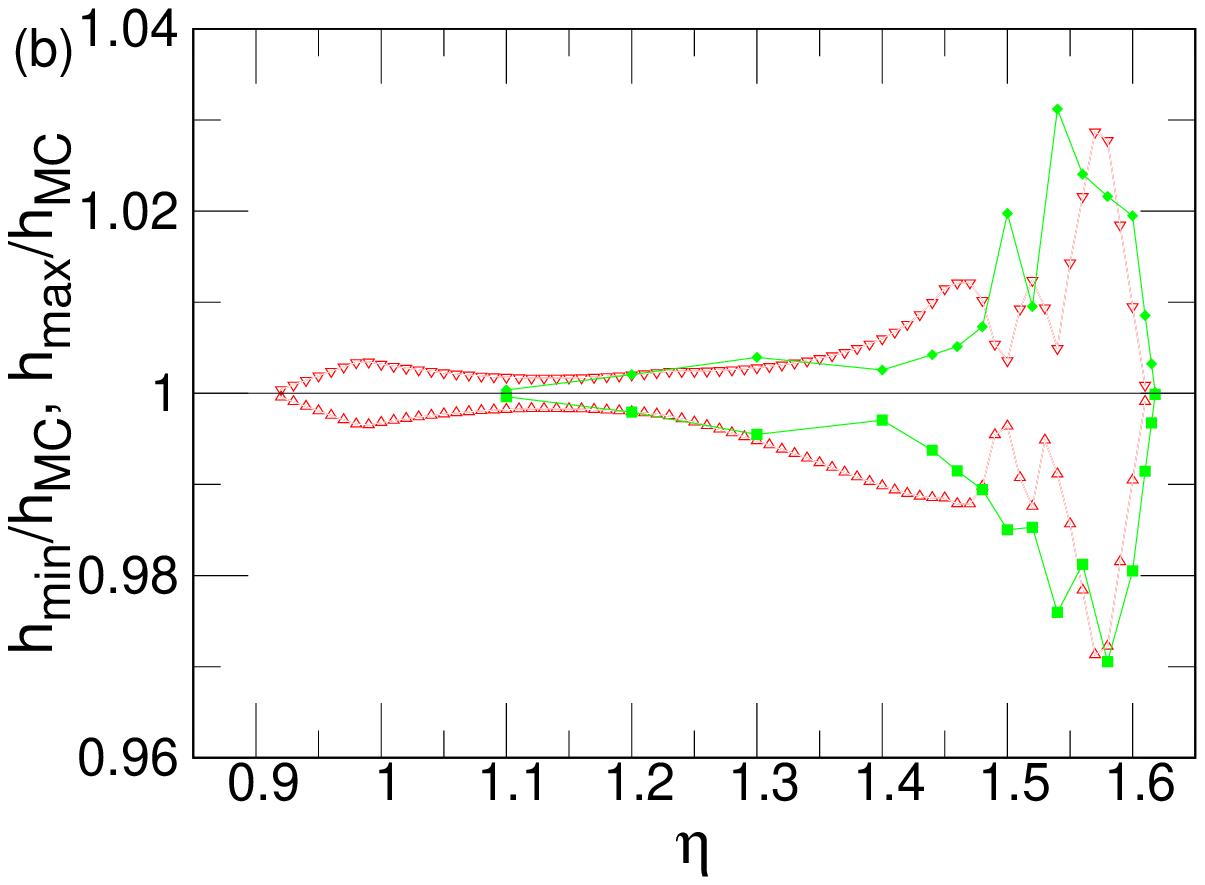}
  \end{center}
 \end{minipage}
 \caption{The $\eta$-dependence of the magnetic fields before and
after the Maxwell construction ($h_{\rm{min}}$, $h_{\rm{max}}$, and $h_{\rm{MC}}$)
for the appearing target jump,
where the upper-left inset of panel (a) illustrates
$h_{\rm{min}}$, $h_{\rm{max}}$, and $h_{\rm{MC}}$.
Main panel (a) shows the $\eta$-dependence of
$h_{\rm{min}}/J_{1}$ by open triangles and closed squares
and $h_{\rm{max}}/J_{1}$ by open inverted triangles and closed diamonds;
the lower-right inset shows that of $h_{\rm{MC}}/J_{1}$
by closed circles and crosses.
Green and red symbols represent results for $N=45$ and $36$ systems,
respectively.
Panel (b) shows the $\eta$-dependence of $h_{\rm{min}}/h_{\rm{MC}}$ and
$h_{\rm{max}}/h_{\rm{MC}}$.
}
 \label{jump5}
\end{figure}
To clarify the range of $\eta$ where the magnetization jump discussed
in the previous paragraph,
let us present the $\eta$-dependence of magnetic fields at the jump
before and after the Maxwell construction,
where the fields are denoted by $h_{\rm{min}}$, $h_{\rm{max}}$, and $h_{\rm{MC}}$
illustrated in the upper-left inset of Fig.~\ref{jump5}(a);
results are depicted in Fig.~\ref{jump5}(a).
When we focus our attention on the edges of this range,
there appears the jump
in the region of $\eta$
up to $\eta=1.618$ for the $N=45$ and $\eta=1.611$ for the $N=36$.
The size difference of this value of $\eta$ is quite small.
On the other hand,
the jump appears in the region down to
$\eta\sim 1.1$ for the $N=45$ and $\eta=0.92$ for the $N=36$;
the size dependence is relatively larger than the other edge.

Next, let us observe the behavior inside the range in Fig.~\ref{jump5}.
One can find a change around $\eta \sim 1.5$
concerning the moving of these fields characterizing the jump
$h_{\rm{min}}/J_{1}$, $h_{\rm{max}}/J_{1}$, and $h_{\rm{MC}}/J_{1}$.
Note markedly that
there is no significant difference in $\eta \sim 1.5$ of the change
among $h_{\rm{min}}/J_{1}$, $h_{\rm{max}}/J_{1}$, and $h_{\rm{MC}}/J_{1}$.
In particular, $h_{\rm{MC}}/J_{1}$ clearly shows the difference
between the regions $\eta < 1.5$ and $\eta > 1.5$, namely,
the gradient in $\eta < 1.5$ differs from that in $\eta > 1.5$.
To clarify the differenece, let us observe $h_{\rm{max}}/h_{\rm{MC}}$
and $h_{\rm{min}}/h_{\rm{MC}}$ show in Fig.~\ref{jump5}(b).
These behaviors observed in Fig.~\ref{jump5} suggest that
the properties of the magnetization jump are different
between $\eta > 1.5$ and $\eta < 1.5$.
It is reasonable that the change of $h_{\rm{min}}$ around $\eta \sim 1.5$
comes from the appearance of the new states
between the jump and the $m=7/9$ plateau; however,
there also appears the change of $h_{\rm{max}}$ almost at the same $\eta$,
which suggests that the properties of the magnetization jump change.
It is still unclear at the present why the change of the jump appears
only in the present model; the reason should be studied in future studies.


\begin{figure}[tbp]
 \begin{minipage}{0.5\hsize}
  \begin{center}
    \includegraphics[width=70mm]{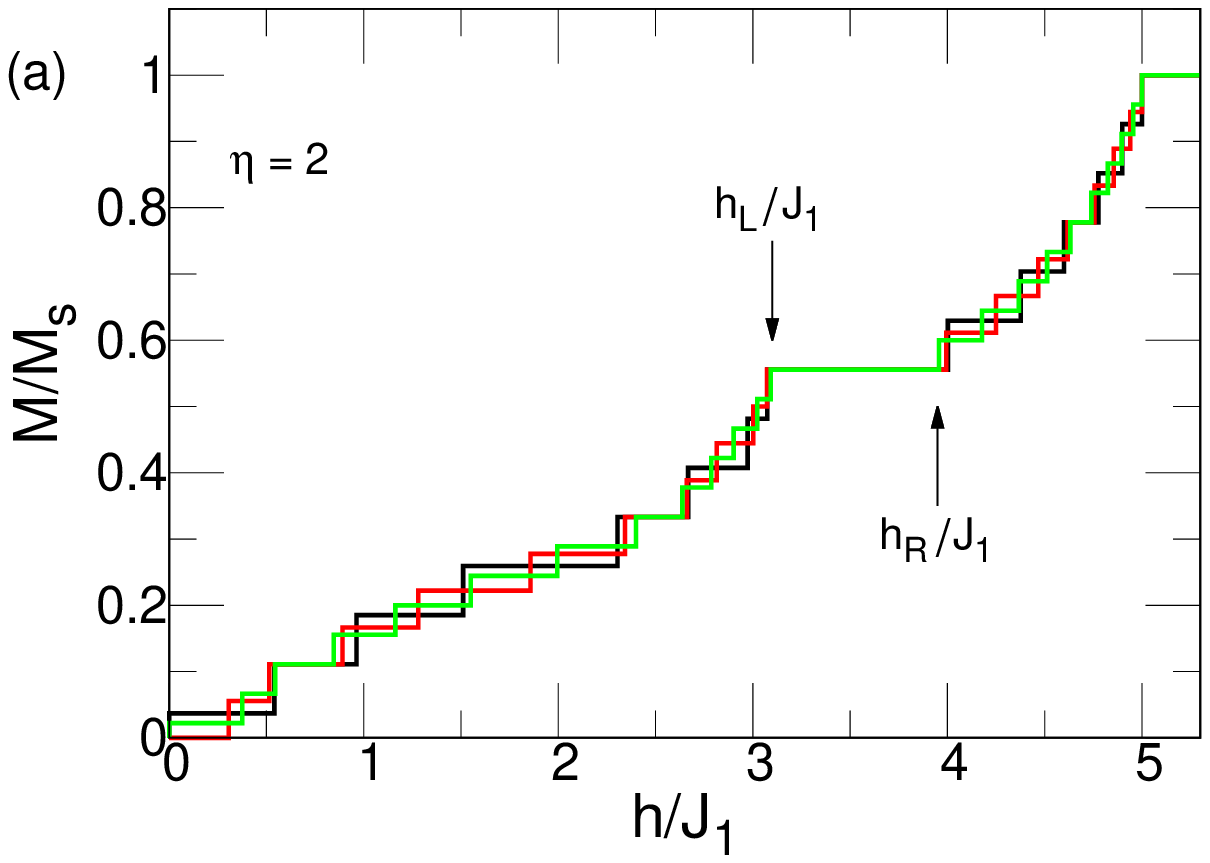}
  \end{center}
 \end{minipage}
 \begin{minipage}{0.5\hsize}
  \begin{center}
    \includegraphics[width=70mm]{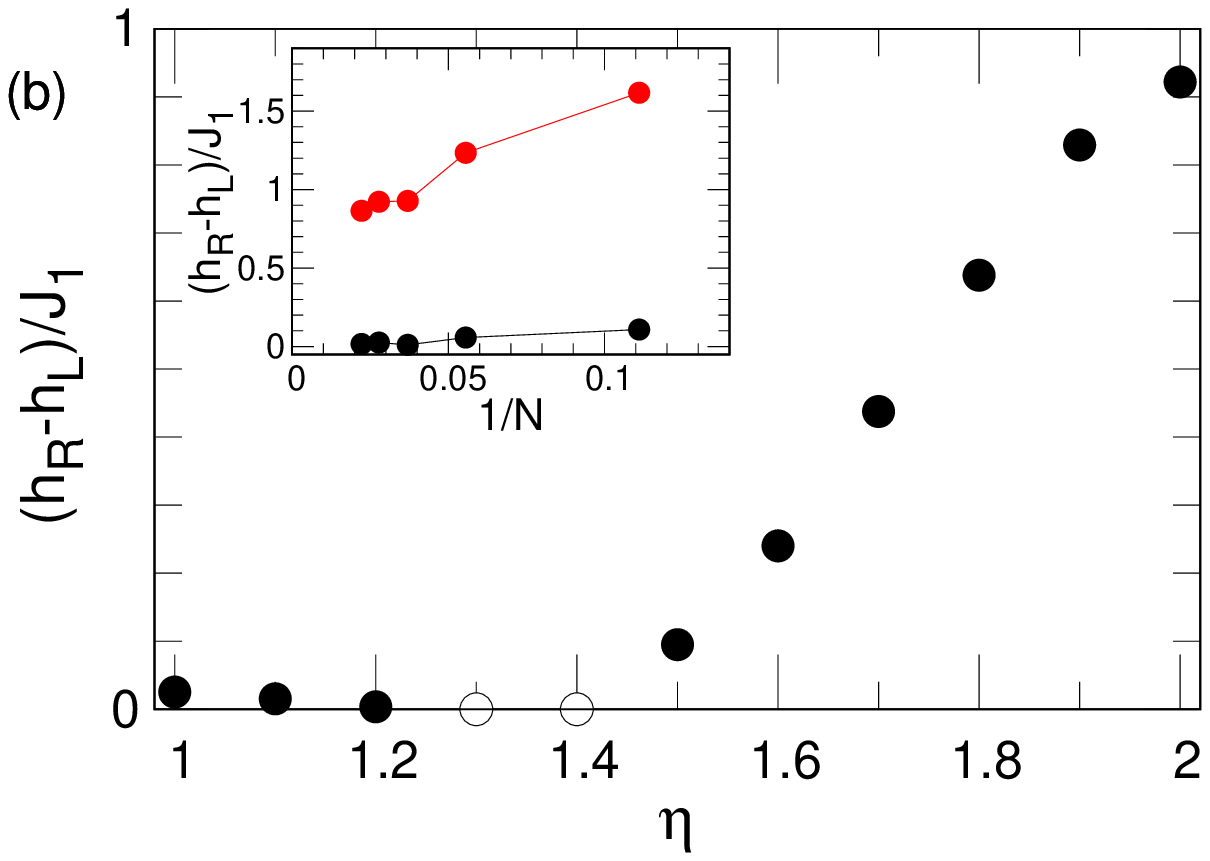}
  \end{center}
 \end{minipage}
 \caption{
Panel (a) depicts the magnetization process for $\eta=2$
where the green, red, and black lines represent results
for $N=45$, $36$, and $27$ systems respectively.
In the main panel (b), $\eta$-dependence of the width of $m=5/9$
in the magnetization process of $N=36$ cluster is depicted.
Closed and open circles correspond to the cases
when the $m=5/9$ states to be realized and unrealized
under a specific magnetic field, respectively.
Namely, the cases of the open circles indicate the appearance
of the magnetization jump.
Inset of panel (b) depicts size-dependence of the $m=5/9$ widths.
The red (black) lines and symbols represent
the results for $\eta=1$ ($\eta=2$).
}
\label{plateau}
\end{figure}
\section{Appearance of the $m=5/9$ plateau}
Let us review the appearance of plateaux
in the case of all the interactions are equivalent, namely $\eta=1$.
Reference~\ref{Furuchi_JPCO2021} reported that for $\eta=1$,
no indication is detected for the plateau at $m=5/9$
although there appear  plateaux at $m=7/9$, 1/3, and 1/9.

Now, let us consider the behavior at $m=5/9$
when $\eta$ is varied.
First, we present the magnetization process for $\eta=2$;
results for $N=45$, 36, and 27 are depicted in Fig.~\ref{plateau}(a).
One easily finds an existing plateau at $m=5/9$.
We define $h_{\rm L}$ ($h_{\rm R}$)
as the lower-field (higher-field) edge of
$m=5/9$.
In the following, let us examine the width of
$m=5/9$
namely,
$h_{\rm L} - h_{\rm R}$.

In the inset of Fig.~\ref{plateau}(b), next,
the $N$-dependence of this width for $\eta=2$
together with the case of $\eta=1$.
Although the width for $\eta=2$ gradually decreases as $N$ is increased,
an extrapolated value of the width to the thermodynamic limit
seems nonzero,
which suggests the $m=5/9$ plateau certainly opens.
For $\eta=1$, in contrast,
finite-size widths are much smaller than those for $\eta=2$;
an extrapolated value of these finite-size results seems to vanish.
These different situations from $\eta=2$ and 1
suggest that the $m=5/9$ plateau closes
in a value of intermediate $\eta$
when $\eta$ is decreased from $\eta=2$.

In order to observe the behavior,
let us observe results of the width for $N=36$
between $\eta=1$ and 2 in the main panel of Fig.~\ref{plateau}(b).
With decreasing $\eta$, the width gradually deceases
down to $\eta\sim 1.5$.
At $\eta\sim 1.4$, finally, the
case of $m=5/9$
encounters the magnetization jump moving from higher $m$
and 
becomes
skipped in the magnetization process.
Below $\eta\sim 1.2$, the
case of
$m=5/9$ recovers its width;
however, the width still shows a small value due to a finite-size effect.
It is noticeable that
the magnetization plateau at $m=5/9$ appears for large $\eta$ whereas
it disappears for small $\eta$
and that
the situation of $m=5/9$ is clearly different from those of
$m=7/9$ and 1/3 reported in Ref.~\ref{Furuchi_JPCO2021}.

\section{Summary}
We have studied the $S=1/2$ Heisenberg antiferromagnet
on the floret-pentagonal lattice
by using the numerical diagonalization method.
The
model
is controlled by the ratio of two interactions,
each of which is determined by the different coordination numbers
of spin sites.
Our numerical-diagonalization calculations up to the 45-site cluster
have clarified the behavior around the five-ninth
of the saturation magnetization.
Further investigations concerning the system on various pentagonal lattices
will contribute much
to our understanding frustration effects in magnetic materials.

\section*{Acknowledgments}

This work was partly supported
by JSPS KAKENHI Grant Numbers
16K05419, 16H01080(J-Physics), 18H04330(J-Physics),
JP20K03866, and JP20H05274.
Nonhybrid thread-parallel calculations
in numerical diagonalizations were based on TITPACK version 2
coded by H. Nishimori.
In this research,
we used the computational resources
of the supercomputer Fugaku provided by RIKEN
through the HPCI System Research projects
(Project IDs: hp200173, hp210068, hp210127, hp210201, and hp220043).
Some of the computations were
performed using facilities of
the Institute for Solid State Physics, The University of Tokyo.


\end{document}